# Canonical Proper Time Formulation for Physical Systems

Tepper Gill
Department of Electrical Engineering and Computational Physics Laboratory
Howard University

James Lindesay
Computational Physics Laboratory
Howard University

**Abstract**

The canonical proper time formulation of relativistic dynamics provides a framework from which one can describe the dynamics of classical and quantum systems using the clock of those very systems. The framework utilizes a canonical transformation on the time variable that is used to describe the dynamics, and does not transform other dynamical variables such as momenta or positions. This means that the time scales of the dynamics are described in terms of the natural local time coordinates, which is the most meaningful parameterization of phenomena such as the approach to equilibrium, or the back reaction of interacting systems. We summarize the formalism of the canonical proper time framework, and provide example calculations of the eigenvalues of the hydrogen atom and near horizon description of a scalar field near a Schwarzschild black hole.

**Outline**

I. General Formulation of Canonical Proper Time framework

II. Alternative formulations of conjugate energy functional

III. An example in quantum mechanics

IV. Usefulness in gravitational physics

V. Conclusion

## I. General Formulation

The general formulation of the canonical proper time transformation has been documented in several references [1,2,3,4]. The key concept can be directly seen by examining the formulation of classical mechanics in terms of Hamilton's equations. The usual definition of the Poisson bracket is given by

$$\{A(p,q), B(p,q)\} \equiv \frac{\partial A}{\partial p}\frac{\partial B}{\partial q} - \frac{\partial A}{\partial q}\frac{\partial B}{\partial p}.$$

1. 1

The time evolution of dynamical parameters is then described in terms of Poisson brackets of that dynamical parameter with the Hamiltonian of the system:

$$\frac{dW(q,p,t)}{dt} = \{H, W(q,p,t)\} + \left(\frac{\partial W}{\partial t}\right)_{q,p}.$$

1. 2

Rather than describe the dynamics using the inertial time t, we will choose to perform a canonical transformation to the proper time of the system. The inertial time t is related to the proper time $\tau$ using the standard Lorentz factor, which can alternatively be expressed in terms of the relationship of the energy of the system H to its rest energy $Mc^2$:

$$dt = \gamma d\tau = \frac{\overset{\bullet}{H}}{Mc^2} d\tau.$$

1. 3

We define a canonical proper energy form that will generate changes in dynamical parameters with respect to the proper time of the system:

$$\frac{dW}{d\tau} = \frac{dW}{dt}\frac{dt}{d\tau} \equiv \{K, W\} + \left(\frac{\partial W}{\partial \tau}\right)_{q,p}.$$

1. 4

Using equations 1. 3 and 1. 4 one can determine the form of the relation between the canonical proper energy and the Hamiltonian. We will require that the canonical proper energy must additionally coincide with the usual Hamiltonian when the proper time coincides with the inertial time:

$$\{K,W\} = \frac{H}{Mc^2}\{H,W\}$$

$$K\big|_{H=Mc^2} = H = Mc^2$$

1. 5

These equations can be formally solved as follows. If $M_o$ is a well defined mass point of the system, then generally the canonical proper energy is given by

$$K = M_o + \int_{M_o}^{H} \frac{H'}{M'} dH'.$$

1. 6

The usual form of Hamilton equations for the Hamiltonian

$$\frac{dq}{dt} = \frac{\partial H}{\partial p} \qquad \frac{dp}{dt} = -\frac{\partial H}{\partial q}$$

1. 7

then implies the form for the generation of the dynamics of $q$ and $p$ variables in terms of the canonical proper energy given by

$$\frac{dq}{d\tau} = \frac{\partial K}{\partial p} \quad , \quad \frac{dp}{d\tau} = -\frac{\partial K}{\partial q}.$$

1. 8

These relations clearly demonstrate the assertion of the canonical proper energy as generating the dynamics in terms of the proper time of the system.

## II. Alternate Formulations of Conjugate Functional K

The manner in which the integration in equation 1. 6 is performed results in various different formulations that can be used to construct the canonical proper energy *K*. The formulation that the authors have found most useful for present purposes is that given by Formulation 1.

Formulation 1:

During the integration, hold the "rest energy" M fixed, and allow Lorentz frame velocity **v** to vary. This results in a form for the canonical proper energy as follows:

$$K_{(1)}[H] = \frac{H^2}{2M} + \frac{M}{2}.$$

2. 1

This formulation is particularly useful for eliminating a square root in the Hamiltonian when developing dynamical equations using $K$. For example, consider the Hamiltonian of a free system of mass M, $H = \sqrt{p^2 + M^2}$. The canonical proper energy $K_{(1)}$ then takes the form

$$K_{(1)} = \frac{p^2}{2M} + M$$

2. 2

This particular example has several noteworthy points of interest:

a. This functional form for the canonical proper energy $K$ is the same as that of the nonrelativistic energy of the system, despite the system being completely relativistic!

b. The momentum parameter in the functional form for $K$ is the momentum as measured in the observer's inertial frame, not the system's proper frame which would measure zero momentum. This clearly demonstrates that this is NOT a Lorentz transformation of the dynamical parameters of the system!

c. If the particle interacts with external influences, the proper frame is NOT inertial, but the proper time is well defined, and is the only true time relevant to the system itself.

d. The troublesome square root in $H$ doesn't appear in $K$.

As an aside, one might note the similarity between the forms obtained in this formulation and those obtained using light cone coordinates [10]. In a light cone formulation, the time and one spatial direction are combined to provide coordinates on the light cone:

$$x^\pm \equiv \frac{1}{\sqrt{2}}(x^0 \pm x^1) \quad, \quad x^j \text{ with } j = 2,3$$

$$\vec{x} \bullet \vec{x} = -x^+ x^- - x^- x^+ + \sum_{j=2}^{3}(x^j)^2$$

2. 3

One typically chooses $x^+$ to play the role of time, and the canonically conjugate parameter $P^-$ to play the role of energy. The longitudinal variable $x^-$ and the transverse variables $x^j$ are like spatial variables. For instance, consider light cone quantization in string theory. A string is parameterized by an affine parameter σ, and is given coordinates x(σ,τ)

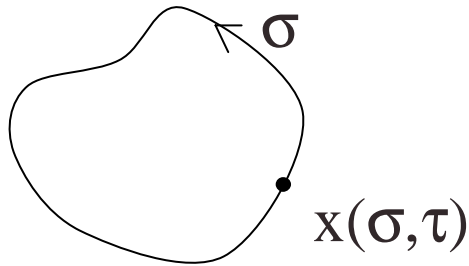

Coordinates of closed string

One can choose $x^+(\sigma, \tau) = \tau$ to characterize the "time" coordinate, and $x^-$, $x^j$ to be the "space" coordinates. The Hamiltonian $P^- = H$ will generate $\tau$ translations. The equation for the mass of the string in terms of its four-momentum squared is a constraint, and the

form of the energy conjugate to $\tau$ can be ascertained, along with the usual quantization conditions:

$$-2P^-P^+ + (P^j)^2 + M^2 = 0 \quad \text{constraint}$$

$$\Rightarrow H = P^- = \frac{(P^j)^2 + M^2}{2P^+}$$

$$[\hat{P}^j, \hat{x}^k] = \frac{\hbar}{i}\delta^{jk}$$

$$[\hat{P}^+, \hat{x}^-] = \frac{\hbar}{i}$$

$$\hat{H}|P^+, P^j\rangle = \frac{(P^j)^2 + M^2}{2P^+}|P^+, P^j\rangle$$

**2. 4**

In these coordinates, $P^+$ acts like another transverse momentum coordinate, and one can see the similarity between this form and the canonical proper energy in formulation 1. It is non-trivial to manifestly display rotational invariance properties using light-cone coordinates, due to the nature of the coordinate identification in equation **2. 3**. One can see that this is not a concern when using the canonical proper time formulation.

There are of course alternative formulations of the canonical proper energy.
Formulation 2:

During the integration in equation 1. 6 hold the three-momentum $P=P_o$ fixed, and allow $H$, $M$ and the frame velocity $v$ to vary:

$$K_{(2)}[H] = (H^2 - P_o^2)^{1/2} = M$$

**2. 5**

This formulation represents the proper energy functional as the invariant mass of the system. The integration serves to build up the mass and velocity of the system from an initial mass point $M_o$ at fixed momentum $P_o$.

Formulation 3:

During the integration in equation 1.6 hold the Lorentz frame velocity $v$ fixed, and allow $H, M$ and the momentum $P$ to vary:

$$K_{(3)}[H] = \frac{H^2}{M}$$

2.6

The integration builds up the mass of the system from an initial mass point $M_o$. This was the formulation first discovered by Gill [9].

One is next led to examine the inclusion of interactions in a canonical proper time formulation. As a concrete example, consider an abelian local gauge symmetry (like electromagnetic interactions). The Hamiltonian generator generally is not a gauge invariant, however one is led to use the equivalent relation

$$dt = \frac{H - V}{H^* - V^*} d\tau$$

2.7

where we define the invariant quantity M in the zero-velocity frame as follows:

$$Mc^2 \equiv H^* - V^*.$$

2.8

This mass is then expected to possess both Lorentz and gauge invariance.

### III. Canonical Proper Time in Quantum Mechanics

To examine physical interpretations of the canonical proper time formulation, we asked ourselves several fundamental questions:

- Is the photon emitted during a quantum transition emitted by the inertial field, or by the recoiling charge?
- Radiation reaction is the interactive response of a radiating charge with the photon DURING the emission process. What does this mean?
- At the most fundamental level, one expects the simplest form of the expression of an emission to be given in terms of the proper frame of the emitter. However, energy conservation is best expressed in the inertial frame of the observer. How is this reconciled?

To examine these and other questions, we performed calculations of the eigenvalues of the canonical proper energy operator for a hydrogen atom to provide further interpretation into the nature of this operator. Calculations involving radiation reaction and other phenomena can be found in the literature [5,6]. The foundations required for developing a scattering formalism using a canonical proper time formulation have also been developed [7,8].

The form of the canonical proper energy for the electron in a hydrogen atom will be obtained using

$$dK = \frac{H - V}{Mc^2} dH$$

$$K = H^* + \int_{H^*}^{H} \frac{H - V}{Mc^2} dH$$

3. 1

Substituting equation 2. 8 into equation 3. 1 gives the relation

$$K = \frac{H^2}{2Mc^2} + \frac{Mc^2}{2} - \frac{(-e\phi^*)^2}{2Mc^2} - \int_{Mc^2 + (-e\phi^*)}^{H} \frac{(-e\phi)dH'}{Mc^2}.$$

3. 2

One next should fix the meaning of the integral in equation 3.2. Using Formulation 1 from Section II above, we obtain H and V using a boost from the * frame (as defined in equation 2.8) as follows:

$$\int_{H^*}^{H} \frac{V'dH'}{H_o^*} = \int_{1}^{\gamma} \frac{\gamma'V^* d\gamma'H^*}{H_o^*} = \frac{\beta^2}{2Mc^2} VH$$

3.3

If we take expectation values of the canonical proper energy relative to the hydrogen atomic states, and examine the lowest order differences between this operator and the Hamiltonian, we can group the various terms of equation 3.2 and compare the expectation values:

$$\langle H \rangle_{n\ell} \equiv Mc^2 + E_{n\ell}$$

$$\langle K \rangle_{n\ell} \cong Mc^2 + E_{n\ell} + \frac{\langle (E_{n\ell} + e\phi)^2 \rangle_{n\ell}}{2Mc^2}$$

3.4

Recall the form of the velocity correction to the fine structure of the hydrogen atom spectrum:

$$\left((pc)^2 + (Mc^2)^2\right)^{1/2} \approx Mc^2 + \frac{p^2}{2M} - \frac{p^4}{8M^3c^2} + \cdots$$

$$\approx Mc^2 + \frac{p^2}{2M} - \frac{(H_o + e\phi)^2}{2Mc^2}$$

3.5

We can immediately conclude that the canonical proper energy K behaves analogously to the Hamiltonian H exclusive of the velocity correction to lowest order. A detailed analysis of this problem is in progress.

## IV. Usefulness of Canonical Proper Time in Gravitational Physics

We next turn our attention to proper time in the vicinity of gravitating objects. In general relativity, the background space-time is influenced by the local energy-momentum density in accordance with the Einstein equation

$$G_{\mu\nu} \equiv R_{\mu\nu} - \frac{1}{2}g_{\mu\nu}R = -\frac{8\pi G}{c^4}T_{\mu\nu}.$$

4. 1

The quantities on the left hand side of the equation are calculated in terms of the geometry, which can be determined by the metric in the space-time interval

$$ds^2 = g_{\mu\nu}dx^\mu dx^\nu.$$

4. 2

One often describes the behavior of the matter-energy fields in terms of an action

$$Action \quad W = \int L\, d^4x,$$

4. 3

where $L$ is a scalar density that has the form $L = \sqrt{-g}\, \tilde{L}$ in terms of the scalar Lagrangian $\tilde{L}$, which is the form usually specified in a physical model, and where $g$ represents the determinant of the matrix formed by the metric $g_{\mu\nu}$. The factor involving the square root of $-g$ insures that the Lagrangian $\tilde{L}$ is a scalar under coordinate transformations, since this will appropriately cancel out the Jacobian factor which arises in an integrand under such coordinate transformations. The energy-momentum tensor is defined in terms of the behavior of the action under variations in the form of the metric

$$\delta W \equiv \frac{1}{2}\int d^4x \sqrt{-g}\, T^{\mu\nu}\delta g_{\mu\nu} \Rightarrow T^{\mu\nu} = \frac{2}{\sqrt{-g}}\frac{\delta W}{\delta g_{\mu\nu}}.$$

4. 4

An energy-momentum tensor defined in this way will be covariantly conserved so that

$$T^{\mu\nu}{}_{;\nu} = 0 = \frac{1}{\sqrt{-g}} \frac{\partial}{\partial x^\nu} \left( \sqrt{-g}\, T^{\mu\nu} \right) + \Gamma^\mu_{\beta\nu} T^{\beta\nu}.$$

4. 5

However, this tensor is not conserved in the flat space-time sense due to the exchange of energy-momentum with the effects of gravitation in the metric and connections.

The Hamiltonian density form is defined in terms of the Legendre transform from the Lagrangian scalar density

$$\Pi = \frac{\partial \mathsf{L}}{\partial(\partial_o \psi)} \quad , \quad \mathsf{H} = \Pi \partial_o \psi - \mathsf{L} = \sqrt{-g}\, T_0{}^0$$

4. 6

The Hamiltonian for a given region of space is obtained by integrating the Hamiltonian density over that region of space (and analogously the Lagrangian is obtained from the Lagrangian scalar density):

$$H = \int d^3 x\ \mathsf{H}$$
$$L = \int d^3 x\ \mathsf{L}$$

4. 7

which, if the fields are solutions of Euler-Lagrange equations can be shown to satisfy functional variations given by

$$\left.\frac{\delta H}{\delta \psi}\right|_\Pi = -\left.\frac{\delta L}{\delta \psi}\right| = -\dot\Pi$$

$$\left.\frac{\delta H}{\delta \Pi}\right|_\Pi = \dot\psi$$

4. 8

In terms of the Hamiltonian density, these equations can be effectively expressed as follows:

$$\frac{\partial \mathsf{H}}{\partial \psi} = -\dot\Pi \qquad \frac{\partial \mathsf{H}}{\partial \Pi} = \dot\psi$$

4. 9

It is understood in the Hamiltonian density equations of motion that derivatives of gradients of the field are properly handled using integrations by parts, or generally

$$\left. \frac{\partial}{\partial \psi}\left(f(x)\bar{\nabla}\psi\right)\right|_{\dot{\psi}} \doteq -\bar{\nabla}f$$

4. 10

We now develop a proper time formulation of these equations. Recall that the relationship between the coordinate time and the proper time can be expressed in terms of the metric relation

$$-ds^2 = d\tau^2 = dt^2\left(-g_{\mu\nu}\frac{dx^\mu}{dt}\frac{dx^\nu}{dt}\right).$$

4. 11

This form immediately gives the previous special relativistic results when the metric form becomes Minkowski space-time, $d\tau^2 = dt^2(1-|v|^2)$. As an example for present consideration, examine Schwarzschild geometry, which describes a static, spherically symmetric space-time outside of the source distribution:

$$ds^2 = -\left(1-\frac{2GM}{r}\right)dt^2 + \frac{dr^2}{\left(1-\frac{2GM}{r}\right)} + r^2 d\vartheta^2 + r^2 \sin^2\vartheta d\varphi^2 = -d\tau^2$$

4. 12

Far from the center, the Schwarzschild geometry behaves essentially like Minkowski space. However, there will be non-trivial differences between coordinate time and proper time in Schwarzschild geometry even for systems that are stationary with respect to the Schwarzschild coordinates:

$$\text{For } r \gg GM \qquad d\tau \to \sqrt{1 - \left|\frac{d\vec{x}}{dt}\right|^2} dt \Rightarrow \frac{\overset{\bullet}{H}}{H^*} = \gamma$$

$$\text{For stationary system} \quad d\tau = \sqrt{1 - \frac{2GM}{r}} dt \Rightarrow \frac{\overset{\bullet}{H}}{H^*} = \sqrt{\frac{1}{1 - \frac{2GM}{r}}}$$

4. 13

For a stationary system at fixed Schwarzschild radial parameter r there is seen to be a well defined connection relating the proper time and the Schwarzschild time coordinates. Neither the Schwarzschild time nor the proper time at fixed finite radial parameter is the locally inertial time that would be represented by the time parameter in a freely falling frame (unless this is the proper system under investigation). A Schwarzschild time lapse *dt* represents the time interval measured by an asymptotic clock that is inertial only at infinite radial parameter.

We demonstrate the utility of using canonical proper time by examining the near horizon form of the Schwarzschild geometry and using the coordinates of Rindler space. Recalling the form of the Schwarzschild line element in equation 4. 12, we choose a radial coordinate that characterizes the proper distance from the horizon:

$$d\rho = \frac{dr}{\sqrt{1 - \frac{2GM}{c^2 r}}} \approx \sqrt{\frac{2GM}{c^2}} \frac{dr}{\sqrt{r - \frac{2GM}{c^2}}}.$$

4. 14

The Schwarzschild metric then takes the form

$$d\tau^2 = \frac{dt^2}{16 G^2 M^2} - d\rho^2 - r^2(\rho)\left(d\vartheta^2 - \sin^2\vartheta d\varphi^2\right).$$

4. 15

Defining Rindler coordinates as follows, the near horizon Schwarzschild coordinates satisfy a metric form that directly demonstrates the nature of the horizon as a coordinate singularity:

$$\rho \equiv 2\sqrt{2GM(r-2GM)}$$
$$\omega \equiv \frac{t}{4GM}$$
$$d\tau^2 = \rho^2 d\omega^2 - d\rho^2 - |d\vec{x}_\perp|$$

4. 16

The singularity at the horizon is seen to be akin to that due to using polar coordinates in a Cartesian space, signifying a coordinate ambiguity at $\rho=0$.

We examine in particular the behavior of a quantum scalar field in this space-time. The Lagrangian for this field will be given by

$$\mathcal{L} \equiv -\frac{\sqrt{-g}}{2}(g^{\mu\nu}\partial_\mu\psi\partial_\nu\psi + m^2\psi^2).$$

4. 17

The canonical momentum density and Hamiltonian density for this system can be directly calculated:

$$\Pi = \frac{\partial \mathcal{L}}{\partial(\partial_\omega \psi)} = \frac{1}{\rho}\partial_\omega \psi$$
$$\mathcal{H} = \frac{\rho}{2}\left\{\Pi^2 + |\vec{\nabla}\psi|^2 + m^2\psi^2\right\}$$

4. 18

The Hamiltonian density is seen to inherit the behavior of the coordinates near the horizon, in that it vanishes as $\rho \to 0$. Dynamics expressed in this description will manifest phenomena that require an understanding of what would be the very high energy behavior of the system, were it studied in flat space, in order to describe its near horizon behavior. The Hamilton equations can then be used to obtain the equation that describes the dynamics of this scalar field

$$\frac{\partial H}{\partial \psi} = -\dot{\Pi} \qquad \frac{\partial H}{\partial \Pi} = \dot{\psi}$$

$$\ddot{\psi} - \rho(\frac{\partial}{\partial \rho}\rho\frac{\partial \psi}{\partial \rho} + \frac{\partial}{\partial x}\rho\frac{\partial \psi}{\partial x} + \frac{\partial}{\partial y}\rho\frac{\partial \psi}{\partial y}) + \rho^2 m^2 \psi = 0$$

4. 19

where dots signify derivatives with respect to the Rindler time coordinate ω, ie $\dot{\psi} \equiv \frac{\partial \psi}{\partial \omega}$.

One might note that the dynamics in Rindler time becomes increasingly static as the horizon is approached. This Rindler time is just a fixed rescaling of Schwarzschild time as defined in equation 4. 16.

We next develop a near horizon proper time formulation of the scalar field. local proper coordinates $x^*$ using the general coordinate transformation equation

$$dx^\mu = \frac{\partial x^\mu}{\partial x^{*\beta}} dx^{*\beta},$$

4. 20

specialized to the time coordinate

$$\frac{dt}{d\tau} = \frac{\partial t}{\partial \tau} + \frac{\partial t}{\partial x^{*j}} \frac{dx^{*j}}{d\tau}.$$

4. 21

The proper time $x^0 \equiv \tau$ is defined by requiring that $\frac{dx^{*j}}{d\tau} = 0$. Similarly, given

$$P^0 = H = \frac{\partial t}{\partial \tau} H^* + \frac{\partial t}{\partial x^{*j}} P^{*j},$$

4. 22

the local rest frame Hamiltonian H* is defined by requiring $P^{*j} = 0$. These objects are well defined at fixed Schwarzschild radial coordinate. We are able to obtain a shortcut relationship between the Hamiltonian *H* and the canonical proper energy *K*. Recall that previously for Minkowski space-times we could express

$$\dot{\gamma} = \frac{H}{H^*} \Rightarrow K = \frac{H^2}{2M} + \frac{M}{2} = \frac{1}{2}\gamma H + \frac{H}{2\gamma}.$$

4. 23

Analogously, for appropriately defined densities, we can identify

$$\mathsf{K} = \frac{1}{2}\frac{dt}{d\tau}\mathsf{H} + \frac{\mathsf{H}}{2\left(\frac{dt}{d\tau}\right)}.$$

4. 24

This allows us to directly obtain an expression for the canonical proper energy density from the energy density.

To express the equations of motion in terms of the proper time, we will rewrite the previous equations 4. 18 and 4. 19 in terms of the canonical proper energy density. The Hamilton equations to be satisfied are given by

$$\frac{\partial \mathsf{K}}{\partial \psi} = -\frac{\partial \Pi}{\partial \tau} \quad , \quad \frac{\partial \mathsf{K}}{\partial \Pi} = \frac{\partial \psi}{\partial \tau}.$$

4. 25

For the scalar field in near horizon coordinates, one makes the following identification using equation 4. 24 and restoring Schwarzschild time units:

$$\frac{d\omega}{d\tau} = \frac{1}{\rho} \quad , \quad \frac{dt}{d\tau} = \frac{4MG}{\rho}$$

$$\mathsf{K} = \frac{1}{4}\left(1 + (\rho/4MG)^2\right)\left\{\Pi^2 + |\bar{\nabla}\psi|^2 + m^2\psi^2\right\} = \frac{1 + (\rho/4MG)^2}{2(\rho/4MG)}\mathsf{H}$$

4. 26

The canonical proper time Hamiltonian density $\mathsf{K}$ is seen to be finite and well behaved near the horizon, despite the vanishing of the Hamiltonian density $\mathsf{H}$. This representation conveniently describes the behavior of the system in terms of proper distance from the horizon and proper time dynamics, and does not exhibit any coordinate pathologies near the horizon. The Hamilton equations describe the dynamics of this field in terms of proper time evolution equations:

$$\frac{\partial \psi}{\partial \tau} = \frac{1}{2}\left(1 + (\rho/4MG)^2\right)\Pi$$

$$\frac{\partial^2 \psi}{\partial \tau^2} - \frac{1}{2}\left(1 + (\rho/4MG)^2\right)\left\{\frac{\partial}{\partial \rho}\left(\frac{1}{2}\left(1 + (\rho/4MG)^2\right)\frac{\partial \psi}{\partial \rho}\right) + \frac{\partial}{\partial x_\perp}\cdot\left(\frac{1}{2}\left(1 + (\rho/4MG)^2\right)\frac{\partial \psi}{\partial x_\perp}\right)\right\} +$$

$$\left(\frac{1}{2}\left(1 + (\rho/4MG)^2\right)m\right)^2 \psi = 0$$

4. 27

Clearly the dynamics of the field expressed in terms of the proper time τ does not inherit the coordinate pathology of the Rindler space-time near ρ=0. The dynamics of this field expressed in terms of the proper radial distance ρ and the proper time parameter τ would represent the natural parameterizations for measurements by local laboratories using standard measuring rods and clocks near the horizon. We would expect thermodynamic time scales and kinematic expressions to be most naturally expressible using these parameters. The presence of the gravitational influences manifests in the difference in the form of this equation as compared to the usual scalar wave equation for flat-spacetimes.

## V. CONCLUSION

We have demonstrated that the canonical proper time formulation of relativistic dynamics gives equivalent physics to inertial time formulations, but provides differing insights into the mechanism of interactions at a fundamental level, especially with regards to radiation reaction, near horizon physics, and analogous processes in which the description of the dynamics can be most intuitively described using the proper time of the interacting system. Some of the calculational difficulties associated with the analytic structure of quantum operators can be transferred into classical calculations of the proper time of a non-inertial system, eliminating the ambiguities of for instance unwanted square-root operations. The resultant forms of the dynamical equations allow one to often use non-relativistic techniques to solve classical or quantum aspects of a problem in terms of the proper time parameter, opening a broadly established range of calculational techniques that can be used for the solutions of relativistic problems.